# Consensus-Driven Group Recommendation on Sparse Explicit Feedback: A Collaborative Filtering and Choquet–Borda Aggregation Framework

Nguyen Van Anh
Hanoi University of Business and Technology, Vietnam
nguyenvananhhubt8888@gmail.com

Ngo Hoang Huy
Intracom University, Vietnam
ngohoanghuy.intracomuni@gmail.com

Ngo Nguyen Khoi
Pitech lab, Hanoi, Vietnam
ngonguyenkhoi98@gmail.com

Pham Thi Ngoc
Intracom University, Vietnam
phamthingoc.intracomuni@gmail.com

Ngo Mai Bao Khanh
Appen Data Technology, Vietnam
khanhnmb@appen.tech

Nguyen Van Quyen
Hai Phong University, Vietnam
quyennv@dhhp.edu.vn

**ABSTRACT.** Group Recommender Systems (GRS) play an essential role in supporting collective decision-making among users with diverse and potentially conflicting preferences. However, achieving stable intra-group consensus becomes particularly challenging when only sparse userID–itemID–rating data are available and no demographic, contextual, or group-level information exists. This paper proposes a consensus-driven hybrid group recommendation framework that integrates neighborhood-based collaborative filtering with fuzzy aggregation to support agreement, fairness, and robustness under sparsity. A composite similarity measure, CBS (Combined Similarity), is derived from two enhanced similarity metrics introduced in prior work: a geometry-based measure that captures rating-pattern structure, and an uncertainty-aware measure that models belief, evidence, and disagreement in sparse co-rating contexts. This combination provides more stable estimation of missing ratings and supports consensus-oriented neighborhood construction. Candidate items are generated by merging per-user top-N predictions and further enriched using the Borda Count mechanism to mitigate skewed rating distributions and reinforce group-level agreement. Final group ratings are computed using the Choquet integral, which flexibly captures heterogeneous user influence while preserving fairness and supporting consensus formation. Experimental results on real-world datasets with different rating distributions show that the proposed method improves group-level consensus, satisfaction, and fairness, while maintaining a balanced level of novelty. Although the model does not rely on social information, its evaluation using trust-aware novelty measures indicates stable behavior in socially structured environments.

## 1. Introduction

Recommender Systems (RS) have become an essential component of modern digital platforms, enabling personalized content delivery across domains such as e-commerce, entertainment, and social media [1]. While individual recommendation has been extensively studied-with established approaches such as collaborative filtering, content-based models, and hybrid techniques-extending recommendation to groups introduces additional challenges that have not been fully addressed [2], [3].

Group Recommender Systems (GRS) must not only predict individual preferences accurately but also aggregate potentially conflicting interests among group members in a fair and consistent manner. This task becomes more difficult in sparse explicit-feedback scenarios, where ratings are limited, and group-level contextual information is largely unavailable. In such settings, sparse and unevenly distributed ratings hinder both neighborhood construction and preference aggregation, leading to reduced stability and limited group consensus [4], [10], [17].

Depending on the richness of available data, GRS methods can be categorized into four major groups [2], [3]:
(i) rich-context GRS that incorporate demographic or content attributes;
(ii) hybrid-metadata GRS combining ratings with partial contextual information;
(iii) sparse explicit-feedback GRS that rely solely on incomplete rating matrices;
(iv) implicit-feedback GRS based on user interaction logs.

Among these, sparse explicit-feedback GRS is particularly relevant for many real-world applications due to privacy considerations and limited availability of group metadata. However, the absence of auxiliary information demands more robust prediction mechanisms and more effective strategies for aggregating individual preferences into collective decisions [4], [17].

Group preference aggregation is central to GRS, traditionally addressed through methods such as Average, Least Misery, or Borda Count. Although conceptually simple, these methods struggle to capture nonlinear interactions or heterogeneous levels of influence between group members [5], [18]. Recent advances in fuzzy logic offer promising alternatives, with the Choquet integral enabling the modeling of complex dependencies and varying influence among users through fuzzy measures [6]–[9].

In sparse settings, accurate prediction of individual ratings is a prerequisite for effective aggregation. Existing research has improved rating prediction via matrix factorization, embedding models, and regularization-based approaches to address cold-start and latent factor inference [11]. Enhanced similarity measures-including TAJ [12], UASim [13], and NNM [14]-have demonstrated improvements in neighbor selection under uncertainty and sparsity.

Motivated by these observations, this study focuses on group recommendation under sparse explicit-feedback conditions and proposes a hybrid framework that integrates collaborative filtering with fuzzy aggregation. Specifically:

- A composite user similarity, CBS (Combined Similarity), is derived from geometry-based and uncertainty-aware similarity components to support reliable neighborhood construction under sparsity.
- Candidate itemIDs for the group are identified by merging the top-N predicted ratings of individual userIDs and enriching this set using Borda Count to mitigate imbalance and reinforce group-level agreement.
- The Choquet integral is employed to aggregate predicted individual ratings, capturing nonlinear user interactions and supporting fair and consensus-oriented group decisions.

The proposed approach is evaluated using a comprehensive set of rating-based and non-rating metrics, covering prediction accuracy, fairness, and novelty. These measures enable an assessment not only of how individual preferences are estimated, but also of how effectively the method supports group consensus while maintaining a balanced level of novelty in the recommended items.

## 2. Related Work

For convenience of presentation, the following notation is used throughout this paper:

U: The set of userIDs.
I: The set of itemIDs.
R: The rating dataset
$R = \{(u, i, r): u \in U, i \in I, r \in [0, +\infty)\}$.
$U_i$: The set of userIDs who rated itemID $i \in I$, with $U_i \subseteq U$.
$I_u$: The set of itemIDs rated by $u \in U$, with $I_u \subseteq I$.
$I_{u,v} = I_u \cap I_v, u, v \in U$.
SU(u,v): The similarity value between $u$ and $v$, where $u, v \in U$.
$N_u^i(k)$: The k-nearest neighbors of userID $u$ who have rated itemID $i \in I$, selected based on the highest similarity scores SU(u,v).
$r_{u,i}$: The rating score given by userID $u$ to itemID $i$.
$r_{\min}$: The smallest rating value in the dataset, $r_{\min} = \min \{r_{u,i} \mid u \in U, i \in I\}$.
$r_{\max}$: The largest rating value in the dataset, $r_{\max} = \max \{r_{u,i} \mid u \in U, i \in I\}$.
$\bar{r}_u$: The mean rating of userID $u$, $\bar{r}_u = \text{mean}\{r_{u,i} \mid i \in I_u\}$.
$\bar{r}_i$: The mean rating of itemID $i$, $\bar{r}_i = \text{mean}\{r_{u,i} \mid u \in U_i\}$.
$E_{\text{trust}}$: The set of trust edges in a social network, $E_{\text{trust}} = \{(u, v) \mid u, v \in U, u \neq v, (u, v) \text{ is a trust relationship}\}$.
$t_{u,v}$: The trust value associated with the edge $(u, v) \in E_{\text{trust}}$, where $t_{u,v} \in [0,1]$.

### 2.1 Collaborative Filtering in Sparse Settings

Collaborative Filtering (CF) is a foundational technique in recommender systems, which predicts a userID's rating for an item based on ratings from similar users [7], [10], [11]. User similarity is typically computed based on co-rated items [11]. For example, the Cosine similarity is defined as:

$$sim(u_1, u_2) = \frac{\sum_{i \in I_{u_1, u_2}} r_{u_1, i} * r_{u_2, i}}{\sqrt{\sum_{i \in I_{u_1, u_2}} r_{u_1, i}^2} * \sqrt{\sum_{i \in I_{u_1, u_2}} r_{u_2, i}^2}} \quad (1)$$

Given the neighbor set $N_u$, the predicted rating for userID $u$ on itemID $i$ is computed as [7], [8]:

$$\widehat{r_{u,i}} = \bar{r}_u + \frac{\sum_{v \in N_u} sim(u, v) * (r_{v,i} - \bar{r}_v)}{\sum_{v \in N_u} |sim(u, v)|} \quad (2)$$

Cosine similarity considers only the magnitude of ratings and does not take into account how many items are jointly rated. This may lead to unreliable similarity estimates in sparse datasets (e.g., $sim(u_1, u_2) = 1$ when $\mid I_{u_1, u_2} \mid = 1$).

To address this issue, Amer and Nguyen [12] introduced TAJ, a similarity metric combining Jaccard (capturing overlap in co-rated items) with an improved cosine-based geometric component to reduce distance-related distortion. Thus, TAJ accounts for both co-rating coverage and the strength of rating agreement. Formally, for $u_1, u_2 \in U$:

$$TAJ(u_1, u_2) = \begin{cases} \frac{(u_{1,c} \cdot u_{2,c})^2}{\|u_{1,c}\| \|u_{2,c}\|^3} \frac{|I_{u_1,u_2}|}{|I_{u_1} \cup I_{u_2}|}, & \text{if } u_{1,c} \cdot u_{2,c} \geq 0 \land \|u_{1,c}\| \leq \|u_{2,c}\| \\ \frac{(u_{1,c} \cdot u_{2,c})^2}{\|u_{1,c}\|^3 \|u_{2,c}\|} \frac{|I_{u_1,u_2}|}{|I_{u_1} \cup I_{u_2}|}, & \text{if } u_{1,c} \cdot u_{2,c} \geq 0 \land \|u_{1,c}\| > \|u_{2,c}\| \\ \frac{u_{1,c} \cdot u_{2,c}}{\|u_{2,c}\|^2} \frac{|I_{u_1,u_2}|}{|I_{u_1} \cup I_{u_2}|}, & \text{if } u_{1,c} \cdot u_{2,c} < 0 \land \|u_{1,c}\| \leq \|u_{2,c}\| \\ \frac{u_{1,c} \cdot u_{2,c}}{\|u_{1,c}\|^2} \frac{|I_{u_1,u_2}|}{|I_{u_1} \cup I_{u_2}|}, & \text{if } u_{1,c} \cdot u_{2,c} < 0 \land \|u_{1,c}\| > \|u_{2,c}\| \end{cases} \quad (3)$$

where $u_{1,c} = (r_{u_1,i} - \overline{r_{u,1}})_{i \in I_{u_1,u_2}}, u_{2,c} = (r_{u_2,i} - \overline{r_{u,2}})_{i \in I_{u_1,u_2}}$.

Belmessous et al. [13] proposed UASim (Uncertainty-Aware Similarity), based on Dempster–Shafer theory, to model uncertainty in sparse settings. UASim considers the ratio of minimum to maximum ratings on co-rated items and combines a belief term $b_x$ with an uncertainty term $u_x$:

$$UASim(u_1, u_2) = b_x + \beta * u_x, \quad (4)$$

, where

$$r_x = \sum_{i \in I_{u_1,u_2}} \frac{\min(r_{u_1,i}, r_{u_2,i})}{\max(r_{u_1,i}, r_{u_2,i})}, \quad b_x = \frac{r_x}{|I_{u_1,u_2}| + w}, \quad u_x = \frac{w}{|I_{u_1,u_2}| + w} \quad (5)$$

and $w = 2, \beta = 0.5$ are empirical constants.

Compared with TAJ, UASim emphasizes uncertainty modeling, which may benefit applications focusing on diversity or serendipity [13].

More recently, Liang et al. [14] introduced Structured Conformalized Matrix Completion (SCMC), which constructs joint confidence regions for multiple missing ratings within the same column of a rating matrix. Unlike traditional conformal prediction methods that operate independently on each user, SCMC provides group-level uncertainty guarantees. SCMC is model-agnostic and can be combined with various matrix completion approaches—from matrix factorization to deep learning—providing improved coverage and reliability. However, SCMC requires substantial computational and memory resources, which may limit its scalability to large datasets. In comparison, similarity-based approaches such as TAJ [12] and UASim [13] generally offer lower computational cost, making them more suitable for large-scale sparse settings.

**2.2 Fairness in Group Recommendation**

Fairness is widely recognized as an important property in Group Recommender Systems (GRS), since recommendations should not systematically favor a subset of group members at the expense of others [3]. In particular, Giap et al. [7] have already analyzed group fairness in depth, proposing Choquet-based aggregation schemes and introducing fairness-oriented indices to evaluate the balance of satisfaction among users.

In this study, we do not further extend the fairness dimension beyond the formulation in [7]. Instead, the Choquet integral is adopted as the aggregation operator following the configuration in [7], while the main focus of the paper is to improve rating prediction quality and to refine the selection of candidate items before applying the Choquet integral for group rating aggregation.

**2.3 Fuzzy Integrals for Decision Making**

Before applying the Choquet integral for group preference aggregation, it is necessary to determine the group candidate set—the collection of items that are most likely to be selected or favored by the group. This set is constructed by combining individual-level preferences with indicators of group-level consensus.

For each userID $u \in g$, where $g$ denotes the group members, an individual Top-N list is first obtained from the itemIDs with the highest rating or predicted rating. The initial group candidate set is then formed as the union of these individual Top-N lists.

However, in highly sparse datasets exhibiting right-skewed rating distributions—where most observed ratings are high—the collaborative filtering model tends to assign similarly high predicted values to many items. Consequently, selecting candidate items solely based on the union of individual Top-N predictions may offer limited discrimination and may fail to capture users' relative ranking preferences.

To mitigate this issue, the candidate set is expanded by incorporating items with the highest Borda Count (BC) scores. For a group $g$, the BC score of an item $i \in I$ is defined as:

$$BC_g(i) = \sum_{u \in g}(n - \pi_u(i) + 1) \tag{6}$$

, where $\pi_u(i)$ is the rank position of item $i$ within user $u$'s preference ordering over all items.

The BC mechanism transforms continuous ratings into discrete rankings, providing additional discrimination even when many ratings cluster near the upper bound. By integrating both the union of individual Top-N lists and the Top-k BC-ranked items, the resulting candidate set captures both individual preferences and group-level preference tendencies, forming a more robust basis for subsequent aggregation using the Choquet integral.

### 2.4 Fuzzy Integral in Group Recommender Systems

Fuzzy aggregation has been widely explored in group recommender systems as a means to incorporate inter-user dependencies that cannot be expressed through linear weighting schemes. Prior work, particularly [7], [8], applies the Choquet integral to combine individual user ratings into a single group score by modeling interaction effects through a fuzzy measure. This operator supports fairness-oriented criteria by allocating weights not only to individual users but also to all user subsets, thereby capturing reinforcing or compensatory relationships within the group.

In the context of group recommendation, the Choquet integral offers a flexible nonlinear mechanism for aggregating predicted ratings while reflecting the influence structure among group members. Given a group $g = \{u_0, \ldots, u_{|g|-1}\} \subseteq U$ and an item $i \in I$, the Choquet integral is defined as:

$$C_g(i;g) = r_{u_{j_0}} + \sum_{k=1}^{|g|-1} \left(r_{u_{j_k}} - r_{u_{j_{k-1}}}\right) \cdot C\left(\{u_{j_k}, \ldots, u_{j_{|g|-1}}\}; g\right) \tag{7}$$

, where $\{j_0, \ldots, j_{|g|-1}\}$ orders users by increasing rating for item $i$, and $C(\cdot; g)$ is a monotonic fuzzy capacity satisfying:

$$\begin{cases} C(\emptyset; g) = 0, \\ C(g; g) = 1, \\ \forall A, B: A \subseteq B \subseteq U, C(A; g) \leq C(B; g) \end{cases} \tag{8}$$

Extensions such as intuitionistic fuzzy Choquet integrals [8], [9] further enhance robustness under uncertainty, which is beneficial in sparse rating environments.

In this work, the Choquet integral is used directly as the nonlinear aggregation operator following the formulation in [7]. We do not extend fairness modeling beyond prior studies; rather, our focus is on improving rating prediction quality and constructing a more reliable group candidate set prior to aggregation. The Choquet integral serves as the final aggregation mechanism, accounting for interactions among users during group preference formation.

### 2.5. Evaluation Metrics for Collaborative Filtering Prediction and Group Recommendation Quality

Assessing the performance of collaborative filtering (CF) prediction and group recommendation systems (GRS) requires a diverse set of metrics that capture accuracy, fairness, and user experience dimensions [1], [11]. The metrics employed in this study include:

#### 2.5.1 Evaluation Metrics for Predicting Ratings

For clarity, we denote:

$$\widetilde{r_{u,i}} = \begin{cases} r_{u,i}, & i \in I_u \\ \widehat{r_{u,i}}, & i \in I \setminus I_u \end{cases} \tag{9}$$

**Mean Absolute Error (MAE).**

MAE measures the average absolute deviation between predicted and actual ratings [11]:

$$MAE = mean\{|\widehat{r_{u,i}} - r_{u,i}|, u \in g, i \in I\} \tag{10}$$

**Root Mean Squared Error (RMSE).**

RMSE places stronger emphasis on large errors and is defined as [11]:

$$RMSE = \sqrt{mean\{(\widehat{r_{u,i}} - r_{u,i})^2, u \in g, i \in I\}} \tag{11}$$

, where $\widehat{r_{u,i}}$ denotes the predicted rating and $r_{u,i}$ the observed rating.

#### 2.5.2. Group-based Recommendation Metrics (Rating-based Evaluation)

**Group satisfaction (group-pref)**

This metric measures the average satisfaction of group members over the recommended item set $I_r \subset I$:

$$group - pref(g, I_r) = mean\{mean\{\widetilde{r_{u,i}} : u \in g\} : i \in I_r\} \tag{12}$$

**Group MAE (MAE-G)**

MAE-G evaluates the deviation between the aggregated group score and individual ratings:

$$MAE - G(g, I_r; C_q) = mean\{|C_q(i;g) - mean\{\widetilde{r_{u,i}} : u \in g\}| : i \in I_r\} \tag{13}$$

**Group RMSE (RMSE-G)**

RMSE-G emphasizes larger deviations between aggregated and individual ratings:

$$RMSE - G(g, I_r; C_q) = \sqrt{mean\{(C_q(i; g) - mean\{\widetilde{r_{u,i}} : u \in g\})^2 : i \in I_r\}} \quad (14)$$

**Fairness – Jain's Index (fairness-1)**
This index measures the equity of satisfaction across group members [3], [7]:

$$fairness_{Jain}(g, I_r) = \frac{(\sum_{u \in g} s_u)^2}{|g| \sum_{u \in g} s_u^2} \quad (15)$$

**Fairness – Variance-based (fairness-2)**
This metric captures the dispersion of group members' satisfaction [7]:

$$fairness_{var}(g, I_r) = 1 - Var\{s(u; I_r) | u \in g\} \in (-\infty, 1] \quad (16)$$

where

$$s(u; I_r) = \frac{\sum_{i \in I_r} \widetilde{r_{u,i}}}{|I_r|} \quad (17)$$

These metrics assess the accuracy of individual rating predictions and evaluate how group recommendations satisfy members while maintaining intra-group consensus, following the evaluation perspective in [7].

## 3. Proposed Method

The proposed framework follows an eight-step workflow, as illustrated in Fig. 1, and is organized into three main stages. First, missing individual ratings are estimated using neighborhood-based collaborative filtering, where user similarity is computed by the proposed CBS measure and neighbors are selected using KNN or TOPSIS. Second, a group-level candidate item set is constructed by merging individual Top-N lists and enriching them with additional items selected via Borda Count. Third, individual preferences—based on observed or predicted ratings—are aggregated using the Choquet integral to produce final group scores, from which the top-N group recommendations are selected.

Each stage is detailed in the following subsections.

### 3.1. Constructing User–User Similarity for Collaborative Filtering

This stage aims to estimate the missing ratings in the sparse rating matrix by applying a k-nearest neighbor (KNN) collaborative filtering approach. The similarity between users is computed using user–user similarity measures together with a filtering condition based on the number of co-rated items

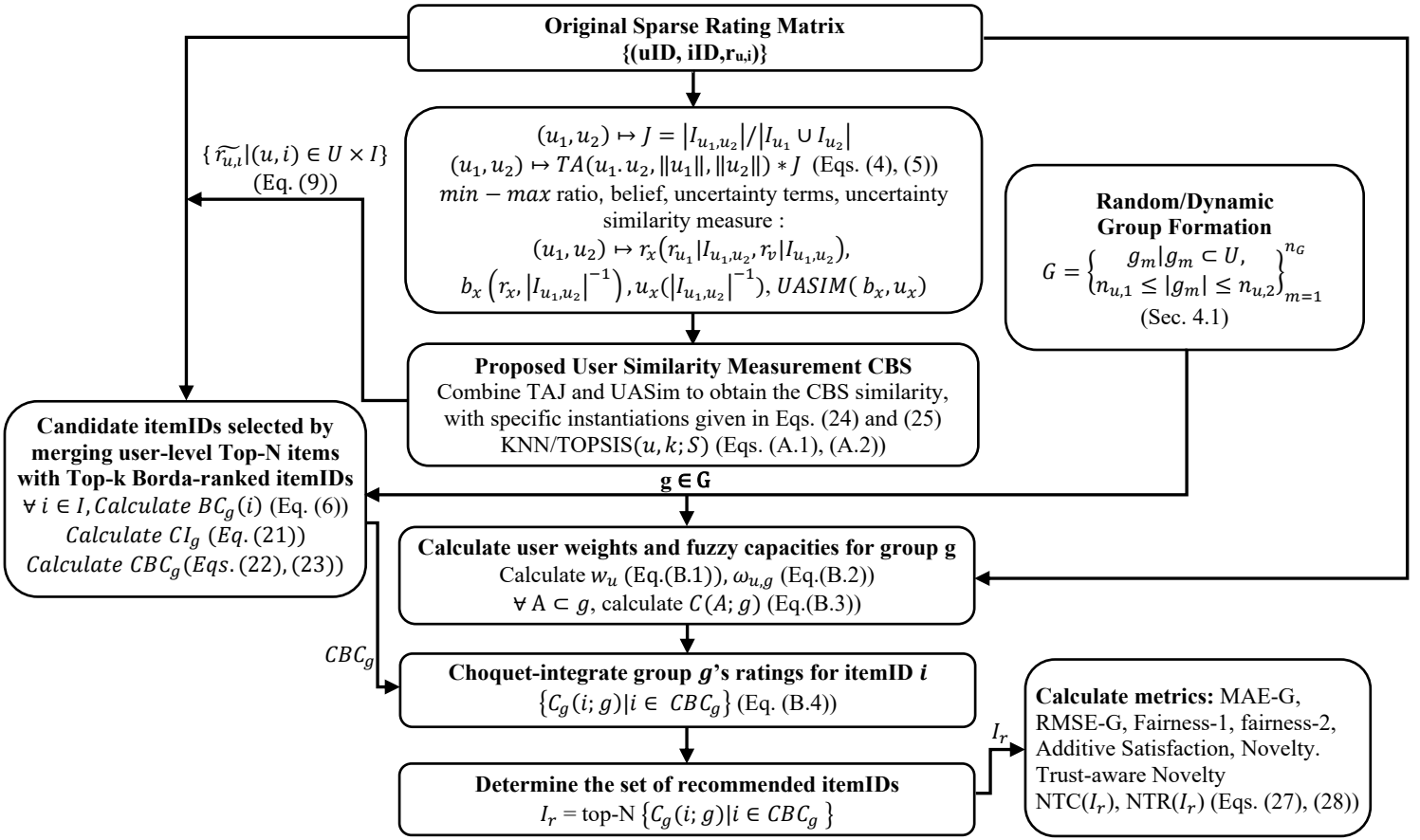

**Fig. 1.** The CBSF-Based Group Recommender System Framework

Fig. 1 illustrates the overall workflow of the proposed GRS model, which proceeds through the following steps.

Starting from the sparse rating matrix $R$, the similarity between two users $u, v \in U$ is first computed using the cosine similarity in Eq. (1). As commonly discussed, Cosine similarity exhibits a limitation that may reduce its reliability in sparse settings: $sim(u_1, u_2) = 1$ whenever $|I_{u_1,u_2}| = 1$.

For example, for the UASIM similarity measure, we define a refined variant, denoted as UASIMJ, by integrating the Jaccard coefficient to capture the overlap between the users' rated item sets:

$$\text{UASIMJ}(u_1, u_2) = \text{UASIM}(u_1, u_2) * \frac{|I_{u_1,u_2}|}{|I_{u_1} \cup I_{u_2}|} \tag{18}$$

In highly sparse rating environments, UASim tends to assign nearly identical similarity values to many user pairs when the number of co-rated items is extremely small. In such cases, the Jaccard component becomes very low while the uncertainty component becomes dominant, causing both belief and disbelief to shrink and reducing the discriminative ability of the similarity function. Consequently, UASim becomes nearly flat under extreme sparsity and is unable to capture the geometric structure of users' rating vectors.

In contrast, the triangle-area similarity (TAJ) preserves geometric characteristics of rating behavior. Even with only one or two co-rated items, TAJ computes the area of the triangle formed by the two rating vectors, capturing both angular deviation and magnitude patterns without being overly dependent on the number of shared ratings. This geometric sensitivity enables TAJ to retain meaningful behavioral distinctions even when UASim becomes heavily influenced by uncertainty.

The combination of TAJ and UASim therefore yields a complementary similarity measure: UASim contributes evidence-aware numerical similarity, whereas TAJ provides geometric stability and preserves rating-pattern structure in highly sparse regions. Experimental results indicate that the linear combination of the two measures performs best when UASim receives a higher weight than TAJ. Assigning UASim as the dominant component maintains essential evidence-driven information, while the contribution of TAJ acts as a corrective geometric counterpart. If the weight on UASim is too low, the combined similarity loses evidence-based discrimination; if it is too high, uncertainty may dominate. A balanced trade-off between expressiveness and robustness in sparse environments is achieved when UASim is maintained as the primary component and TAJ serves as a secondary stabilizing factor.

The predicted rating for user $u$ on item $i$ is then given by:

$$\widehat{r_{u,i}} = \bar{r}_u + \frac{\sum_{v \in N_u} CBS(u, v) * (r_{v,i} - \bar{r}_v)}{\sum_{v \in N_u} |CBS(u, v)|} \tag{19}$$

in which $N_u$ denotes the set of neighbors of user $u$.

Based on these components, we obtain a collaborative filtering method that prioritizes neighbors who share a substantial number of co-rated items.

**Algorithm 1. CBS-Based Collaborative Filtering**

**Input:** A group of users $g \subseteq U$, and a parameter $k$ representing the number of neighbors for each user.
**Output:** A predicted rating matrix for all users in $g$.
**Step 1.**
For each user $u \in U$, randomly split the userID–itemID interactions into training and test subsets (e.g., 80:20), obtaining $\{u, I_{u,train} : u \in U\}$ and $\{u, I_{u,test} : u \in U\}$.
Define

$$I_{\text{train}} = \bigcup_{u \in U} I_{u,\text{train}}, I_{\text{test}} = \bigcup_{u \in U} I_{u,\text{test}} \tag{20}$$

**Step 2. Compute user–user similarity on the training set.**
For every pair of distinct users $u_1, u_2 \in U$:
2.1. Compute the TAJ similarity $TAJ(u_1, u_2)$ using (3), based only on ratings of items $i \in I_{u,\text{train}}$.
2.2. Compute the UASIM similarity $UASIMJ(u_1, u_2)$ using (4)–(5), also restricted to $i \in I_{u,\text{train}}$.
2.3. Combine these measures using (16) to obtain the weighted similarity $CBS(u_1, u_2; I_{\text{train}})$.
This yields a similarity table $\{MS(u_1, u_2) : u_1, u_2 \in U, u_1 \neq u_2\}$.
**Step 3. Predict ratings for users in the group $g$.**
For each user $u \in g$:
For each item $i \in I$:
- Determine the neighbor set $N_u^i(k)$ using KNN or TOPSIS [Eqs. (A.1), (A.2)], with rating information restricted to $I_{u,\text{train}}$, based on (2) and the similarity table $\{MS(u_1, u_2)\}$.
- If $N_u^i(k) \neq \emptyset$, compute the predicted rating $\hat{r}_{u,i}$ using (17).
Otherwise, set $\hat{r}_{u,i} = $ NaN.
Return the prediction matrix $\{\hat{r}_{u,i}\}_{u \in g, i \in I}$.

The computational complexity of Algorithm 1 is $O(|g|*|U|*|I|*\log k)$.

## 3.2 Group rating prediction for candidate items

Given a group of users $g \subseteq U$, the candidate item set $CI(g;N)$ is constructed by taking the union of the top-$N$ itemIDs with the highest rating (or predicted rating) of each userID in the group. For each user $u$ and item $i$, the rating value $\tilde{r}_{u,i}$ is taken from the original data when available, or from the predicted value otherwise. Formally:

$$CI_g = \bigcup_{u \in g} N\text{-top}\{i \in I : \tilde{r}_{u,i} \to \max\} \quad (21)$$

and $N$ is an experimental parameter (e.g., $N = 40$).

To enhance candidate coverage, we additionally select the $k$-top items according to the group Borda Count:

$$BC_g^{(k)} = k\text{-top}\{BC_g(i) : i \in I\} \quad (22)$$

and define the extended candidate set as:

$$CBC_g = CI_g \cup BC_g^{(k)} \quad (23)$$

The set $CBC_g$ serves as the input for the subsequent group-preference aggregation stage.

**Algorithm 2. CBSF: Collaborative–Borda–Symmetric Fuzzy Integral for Group Recommendation**

**Input:** a user group $g \subseteq U$, parameter $N$.
**Output:** a set of $N$ recommended items for group $g$.
1. Compute the candidate item set $CBC_g$ using Equation (23).
2. For each item $i \in CBC_g$:
   – Compute the group-level rating $C_q(i;g)$ using the Choquet integral, as defined in Equation (7) and described in [7].
3. Select the top-$N$ items for group $g$ based on their aggregated scores:
$$RI_g = \text{top-}N\{C_q(i;g) \mid i \in CBC_g\}$$
4. Return: $RI_g$.

The dominant cost of Algorithm 2 comes from computing the Choquet integral for each candidate item, which requires $O(|g|^3)$ operations per item under the formulation in [7]. Since this is applied to $N$ items, the overall complexity is $O(|g|^3 * N)$.

## 4. Experiments

This section presents the experiments conducted to evaluate the effectiveness of the proposed method in improving collaborative filtering rating prediction and group recommendation quality under sparse explicit-feedback conditions. The evaluations are performed on two real-world datasets: MovieLens (MovieLens100K) and FilmTrust. The two datasets represent different rating characteristics—MovieLens with a relatively balanced distribution and FilmTrust with a right-skewed distribution and trust relations among users.

A diverse set of evaluation metrics is employed, including both rating-based and non-rating-based measures. The proposed method is compared against common baselines and previously published approaches, including the method introduced by Hoang Van Giap et al. [7]. For FilmTrust, trust information is not used for model construction but is included in the evaluation to analyze how well the recommendations align with user–user relational structure.

### 4.1. Experimental Datasets (English Version – Final)

The proposed method is evaluated on two highly sparse datasets with distinct statistical characteristics in order to examine its adaptability and stability across different rating distributions.

The MovieLens 100K dataset is used as the primary benchmark. It contains 943 users, 1,682 movies, and 100,000 ratings on a 1–5 scale, with a sparsity level of 93.7%. Although sparse, MovieLens exhibits a relatively balanced distribution of ratings across users and items, allowing the model to capture generalizable preference patterns.

The second dataset, FilmTrust, is derived from the FilmTrust social network, where users provide movie ratings and may also establish trust relationships with others. It includes 1,508 users, 2,071 movies, and 35,497 ratings on a scale of 0.5–4.0, along with an explicit trust matrix. FilmTrust is extremely sparse (98.9%) and displays a right-skewed rating distribution, meaning that users tend to assign disproportionately higher ratings—an instance of the commonly observed positive bias in social-network data.

For evaluating collaborative filtering prediction accuracy, the items rated by each user $u$ are split randomly into training (80%) and test (20%) subsets. The global training set is constructed by aggregating all per-user training subsets.

To simulate group settings, $n_G = 120$ groups with sizes ranging from $n_{u,1} = 3$ to $n_{u,2} = 30$ users are generated at random, reflecting practical scenarios such as dynamic groups [7]. These setups emulate real-world contexts where group membership is flexible rather than predefined.

### 4.2. Experimental Parameters

To evaluate the performance of the proposed model—CBF (Choquet–Borda-based Fuzzy Recommender), a series of experiments are conducted by comparing it against several baseline methods.

The experimental parameters are configured as follows:
- For the MovieLens 100K dataset, the number of neighbors in the KNN-based collaborative filtering module is set to k = 100, while for FilmTrust, this value is set to k = 50.
- The parameter N-filter = 40 specifies the number of items selected for each user based on individual predicted or observed ratings.
- The parameter N-Borda = 50 denotes the number of additional candidate items selected according to their Borda Count ranking during group candidate construction.
- The final number of candidate items is limited to N-top = 40 for MovieLens 100K for FilmTrust.
- Parameters associated with UASIM are kept consistent with the default configuration reported in [13], corresponding to formulas (4) and (6).

These parameter values are chosen through preliminary trials, with adjustments made to account for the scale and statistical characteristics of each dataset, including the structure of the corresponding trust network.

### 4.3. Evaluation Metrics

The evaluation includes two groups of metrics. The first group-{MAE, RMSE}-assesses the predictive accuracy of the collaborative filtering module.

The second group-{MAE-G, RMSE-G, group-pref, fairness-1, fairness-2}—evaluates the quality of group recommendations based on the individual user ratings, which in most cases are predicted ratings. These metrics measure the group's overall satisfaction and the degree of consensus among members.

### 4.4. Experimental Results and Discussion

**Experiment 1: Evaluating the collaborative filtering model with the proposed CBS similarity measure**

The first experiment aims to assess the effectiveness of the KNN-based collaborative filtering model when applying the proposed composite similarity measure CBS. The evaluations are conducted on the MovieLens 100K dataset following the training–testing protocol described in [13].

In each training run, the similarity matrix is computed entirely from the training set to ensure independence from the test set. The data preparation and neighbor-selection procedures consist of two main steps:

**(1) User-based data splitting.**

For each user, all ratings are randomly divided using an 80:20 ratio, ensuring that each user retains at least five ratings in the training set. In addition, every item appearing in the test set must also appear at least once in the training set to allow valid predictions.

**(2) Neighbor-selection methods.**

Two methods are applied to determine user neighbors:
- KNN: selecting up to $k$ users with the highest similarity scores.
- TOPSIS [13]: each pair of users is represented by three criteria — similarity $S$, uncertainty $U$, and dissimilarity $\bar{S} = 1 - S - U$. After normalization, a closeness score $C$ is computed using TOPSIS, and the top–k users with the highest $C$ values are selected as neighbors.

Details of the TOPSIS procedure are provided in Appendix A.

$$CBS(u_1, u_2) = \begin{cases} a_1 * UASIMJ(u_1, u_2) + (1 - a_1) * TAJ(u_1, u_2), & UASIMJ(u_1, u_2) < th_1 \\ UASIMJ(u_1, u_2), & else \end{cases} \quad (24)$$

, where $a_1 = 0.8$, $th_1 = 0.2$.

The similarity matrix is computed for CBS and for baseline similarity measures including Cosine, TAJ, UASim, and UASIMJ, using only training data. Predicted values are generated for all (user, item) pairs in the test set and evaluated using RMSE and MAE.

**Table 1. Comparison between KNN and TOPSIS neighbor selection**

| Methods | RMSE | | MAE | |
|---|---|---|---|---|
| | KNN | TOPSIS | KNN | TOPSIS |
| Cosine [7] | 1.0989 | 1.1172 | 0.8399 | 0.8705 |
| TAJ [12] | 1.0648 | 1.0582 | 0.8095 | 0.8058 |
| UASIM [14] | 1.0853 | 1.1051 | 0.8259 | 0.8537 |
| UASIMJ | 1.0661 | 1.0577 | 0.8104 | 0.8059 |
| **CBS** | 1.0661 | **1.0575** | 0.8102 | **0.8057** |

The results show that CBS and UASIMJ — both incorporating the number of co-rated items between two users — achieve lower prediction errors than Cosine, TAJ, and UASim when combined with TOPSIS. CBS yields the lowest error, slightly outperforming UASIMJ.

TOPSIS enhances the performance of CBS and UASIMJ compared with KNN. Cosine performs less consistently due to limitations in sparse settings. TAJ performs well in several cases but less stably than CBS or UASIM.

Because UASim does not incorporate the extent of overlap in co-rated items, it tends to be less flexible under highly sparse conditions.

Compared with the NNM algorithm [14] (RMSE = 1.0149; MAE = 0.7966), CBS achieves comparable error levels while requiring significantly lower computational cost.

**Table 2. Comparison of prediction time on the test set**

| Methods | Time Complexity | Runtime (s) | Space Complexity | Runtime (MB) |
|---|---|---|---|---|
| Cosine [7] | $O(|U|^2 \times |I|)$ | 262.04 | $O(|U|^2)$ | 18.2 |
| TAJ [12] | $O(|U|^2 \times |I|)$ | 262.04 | $O(|U|^2)$ | 18.2 |
| UASim [14] | $O(|U|^2 \times |I|)$ | 262.04 | $O(|U|^2)$ | 18.2 |
| UASIMJ | $O(|U|^2 \times |I|)$ | 262.04 | $O(|U|^2)$ | 18.2 |
| CBS | $O(|U|^2 \times |I| + |U| \times k \log(k))$ | 424.24 | $O(|U|^2)$ | 36.4 |
| NNM [16] | $O(|U|^2 \times |I|^2)$ | 3614.90 | $O(|U|^2 \times |I|)$ | 3760.1 |

The CF-based methods require substantially less memory and execution time than NNM. Fig. 2 illustrates this difference.

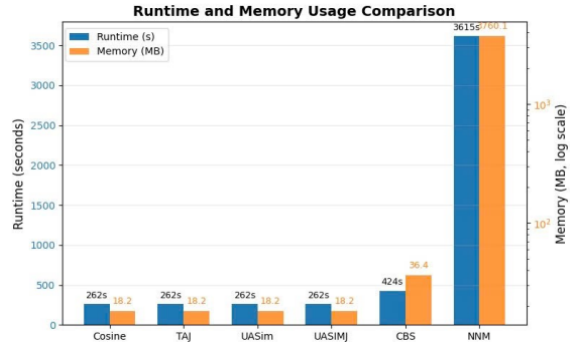

**Fig. 2**. Comparison of Execution Time and Memory Usage between NNM and CBS

Based on Experiment 1, the proposed GRS model adopts the CBS similarity measure combined with TOPSIS-based neighbor selection, as described in Algorithm 1.

**Experiment 2**

In this experiment, the user groups that were previously constructed are used to evaluate group recommendation performance on the MovieLens 100K dataset. The proposed method is assessed using the metrics Additive Satisfaction, fairness-1, fairness-2, MAE-G, and RMSE-G at recommendation list sizes **N = 5, 10, 15, 20, 25, 30, 35, 40**.

**Table 3. Comparison of evaluation metrics**

| N-top | Additive Satisfaction | | RMSE-G | | MAE-G | | fairness-1 | | fairness-2 | |
|---|---|---|---|---|---|---|---|---|---|---|
| | [7] | Proposed | [7] | Proposed | [7] | Proposed | [7] | Proposed | [7] | Proposed |
| 5 | 4.9968 | **4.9976** | 0.0043 | **0.0037** | 0.0031 | **0.0023** | 0.9997 | 0.9997 | 0.9993 | 0.9993 |
| 10 | 4.9982 | 4.9982 | **0.0027** | 0.0031 | 0.0017 | 0.0017 | 0.9998 | 0.9998 | 0.9996 | 0.9996 |
| 15 | 4.9152 | **4.9231** | 0.1096 | **0.0946** | 0.0847 | **0.0768** | 0.9999 | 0.9999 | 0.9954 | 0.9996 |
| 20 | 4.8446 | **4.8739** | 0.2010 | **0.1617** | 0.1553 | **0.1260** | 0.9999 | 0.9999 | 0.9900 | **0.9938** |
| 25 | 4.7945 | **4.8276** | 0.2683 | **0.2231** | 0.2054 | **0.1723** | 0.9993 | **0.9999** | 0.9850 | **0.9898** |
| 30 | 4.7510 | **4.7949** | 0.3251 | **0.2695** | 0.2489 | **0.2050** | 0.9979 | **0.9999** | 0.9803 | **0.9861** |
| 35 | 4.7123 | **4.7674** | 0.3737 | **0.3082** | 0.2876 | **0.2325** | 0.9966 | **0.9998** | 0.9759 | **0.9826** |
| 40 | 4.6777 | **4.7403** | 0.4163 | **0.3440** | 0.3222 | **0.2596** | 0.9950 | **0.9992** | 0.9717 | **0.9794** |

The accompanying figures illustrate the variation tendencies of each metric, providing a visual comparison across methods:

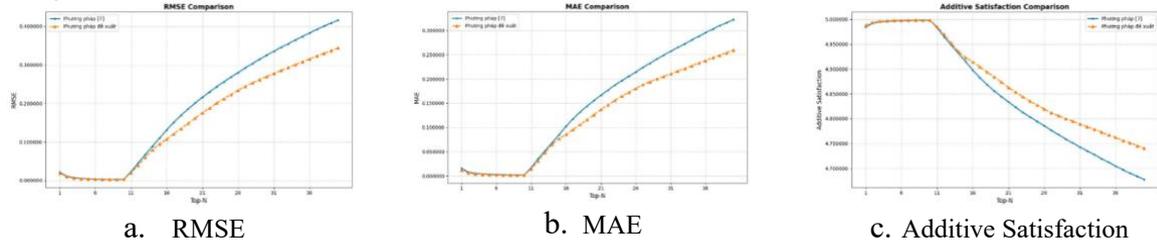

a. RMSE     b. MAE     c. Additive Satisfaction

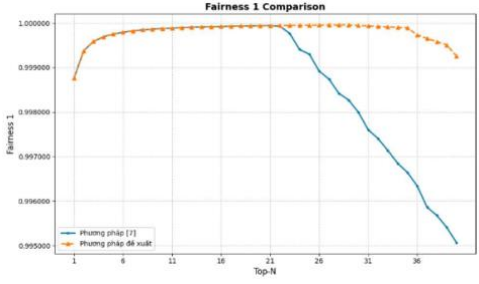
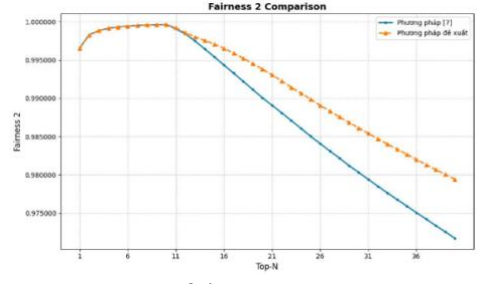

d. fairness-1  e. fairness-2

**Fig. 3.** Group Satisfaction and Consensus Metrics on MovieLens 100K

The results of Experiment 2 show that group recommendation quality improves when incorporating the main components of the proposed framework, including:

(i) replacing the Cosine similarity measure in [7] with the proposed CBS similarity to enhance the quality and stability of individual rating prediction;
(ii) refining the neighbor-selection strategy, where alternatives such as KNN or TOPSIS can be employed depending on the characteristics of the rating distribution and data sparsity;
(iii) enriching the group candidate set by adding items selected via Borda Count prior to aggregation with the Choquet integral.

Together, these components contribute to improved group-level satisfaction, fairness, and consensus across different group sizes.

The metrics RMSE-G, MAE-G, group satisfaction, and fairness (fairness-1, fairness-2) overall demonstrate improvement across multiple group sizes. These findings indicate that the proposed components support improved group recommendation quality under sparse data and heterogeneous preferences, while maintaining group-level consensus.

**Experiment 3.**

In this experiment, the previously introduced group configuration and experimental setup are retained and applied to the FilmTrust dataset. Unlike trust-aware collaborative filtering approaches such as RTCF [19], which directly exploit the trust network in FilmTrust, CBS does not rely on trust information. Even so, the individual prediction results in Table 4 show that CBS attains lower RMSE and MAE than several trust-based variants.

$$CBS(u_1, u_2) = \begin{cases} a_2 * TAJ(u_1, u_2) + (1 - a_2) * UASIM(u_1, u_2), & TAJ(u_1, u_2) < th_2 \\ TAJ(u_1, u_2), & else \end{cases} \quad (25)$$

, where $a_2 = 0.6$, $th_2 = 0.8$.

**Table 4. Comparison between KNN and TOPSIS neighbor selection**

| Methods | RMSE | | MAE | |
|---|---|---|---|---|
| | KNN | TOPSIS | KNN | TOPSIS |
| Cosine [7] | 0.8384 | 0.9086 | 0.6417 | 0.7008 |
| TAJ [12] | 0.8274 | 0.8255 | 0.6306 | 0.6301 |
| UASIM [14] | 0.8225 | 0.8627 | 0.6249 | 0.6655 |
| UASIMJ | 0.8280 | 0.8246 | 0.6312 | 0.6293 |
| **CBS** | 0.8189 | 0.8124 | 0.6261 | 0.6213 |

**Table 5. Comparison between CBS and RTCF on the FilmTrust dataset**

| Methods | RMSE | MAE |
|---|---|---|
| RTCF | 0.879 | 0.667 |
| Yilmaz | 0.912 | 0.685 |
| **CBS** | **0.869** | **0.658** |

To maintain consistency with the evaluation scale used for MovieLens 100K in the subsequent stages, the FilmTrust ratings are normalized to the interval [1, 5]. The complete group-recommendation pipeline is also preserved, consisting of:

(iv) individual rating prediction using the CBS similarity measure
(v) neighbor selection via TOPSIS
(vi) enrichment of the group candidate set using the Borda Count method.

Group ratings are then computed using the Choquet integral.

Group-level evaluation is performed for top-N recommendation lists with $N = 5, 10, 15, 20, 25, 30, 35, 40$ allowing an analysis of how performance varies as the list length increases.

**Table 6. Group satisfaction and fairness measures on FilmTrust for KNN/TOPSIS**

| N-top | Method | | Additive Satisfaction | RMSE-G | MAE-G | fairness-1 | fairness-2 |
|---|---|---|---|---|---|---|---|
| 5 | [7] | KNN | 4.461 | 0.688 | 0.539 | 0.954 | 0.942 |
| | [7]-BC | | 4.482 | 0.684 | 0.518 | 0.949 | 0.936 |
| | CBS | | **4.528** | **0.608** | **0.471** | **0.974** | **0.952** |
| | [7] | TOPSIS | 4.441 | 0.735 | 0.559 | 0.928 | 0.927 |
| | [7]-BC | | 4.441 | 0.734 | 0.559 | 0.928 | 0.928 |
| | CBS | | 4.319 | 0.844 | 0.681 | 0.912 | 0.922 |
| 10 | [7] | KNN | 4.397 | 0.765 | 0.603 | 0.937 | 0.930 |
| | [7]-BC | | 4.423 | 0.762 | 0.577 | 0.929 | 0.919 |
| | CBS | | **4.471** | **0.675** | **0.529** | **0.961** | **0.945** |
| | [7] | TOPSIS | 4.389 | 0.797 | 0.610 | 0.911 | 0.916 |
| | [7]-BC | | 4.389 | 0.796 | 0.611 | 0.911 | 0.916 |
| | CBS | | 4.255 | 0.9182 | 0.745 | 0.884 | 0.908 |
| 15 | [7] | KNN | 4.356 | 0.814 | 0.643 | 0.923 | 0.921 |
| | [7]-BC | | 4.390 | 0.806 | 0.606 | 0.915 | 0.899 |
| | CBS | | **4.438** | **0.715** | **0.562** | **0.950** | **0.939** |
| | [7] | TOPSIS | 4.350 | 0.841 | 0.649 | 0.899 | 0.908 |
| | [7]-BC | | 4.351 | 0.839 | 0.649 | 0.900 | 0.909 |
| | CBS | | 4.215 | 0.965 | 0.784 | 0.867 | 0.898 |
| 20 | [7] | KNN | 4.326 | 0.851 | 0.674 | 0.911 | 0.914 |
| | [7]-BC | | 4.365 | 0.839 | 0.628 | 0.902 | 0.892 |
| | CBS | | **4.411** | **0.748** | **0.589** | **0.941** | **0.933** |
| | [7] | TOPSIS | 4.323 | 0.871 | 0.677 | 0.892 | 0.903 |
| | [7]-BC | | 0.909 | 0.869 | 0.676 | 0.893 | 0.904 |
| | CBS | | 4.188 | 0.996 | 0.809 | 0.855 | 0.891 |
| 25 | [7] | KNN | 4.303 | 0.879 | 0.697 | 0.901 | 0.907 |
| | [7]-BC | | 4.347 | 0.864 | 0.641 | 0.892 | 0.886 |
| | CBS | | **4.389** | **0.776** | **0.611** | **0.933** | **0.927** |
| | [7] | TOPSIS | 4.303 | 0.894 | 0.697 | 0.885 | 0.899 |
| | [7]-BC | | 4.304 | 0.891 | 0.695 | 0.886 | 0.899 |
| | CBS | | 4.168 | 1.016 | 0.827 | 0.847 | 0.886 |
| 30 | [7] | KNN | 4.284 | 0.902 | 0.715 | 0.892 | 0.902 |
| | [7]-BC | | 4.331 | 0.881 | 0.652 | 0.886 | 0.883 |
| | CBS | | **4.369** | **0.801** | **0.631** | **0.924** | **0.922** |
| | [7] | TOPSIS | 4.285 | 0.914 | 0.714 | 0.880 | 0.895 |
| | [7]-BC | | 4.286 | 0.911 | 0.713 | 0.881 | 0.896 |
| | CBS | | 4.149 | 1.035 | 0.843 | 0.839 | 0.881 |
| 35 | [7] | KNN | 4.268 | 0.921 | 0.731 | 0.885 | 0.897 |
| | [7]-BC | | 4.316 | 0.894 | 0.659 | 0.878 | 0.878 |
| | CBS | | **4.353** | **0.820** | **0.647** | **0.917** | **0.918** |
| | [7] | TOPSIS | 4.269 | 0.933 | 0.729 | 0.874 | 0.890 |
| | [7]-BC | | 4.271 | 0.929 | 0.728 | 0.875 | 0.892 |
| | CBS | | 4.136 | 1.046 | 0.852 | 0.835 | 0.879 |
| 40 | [7] | KNN | 4.254 | 0.938 | 0.745 | 0.879 | 0.894 |
| | [7]-BC | | 4.297 | 0.909 | 0.669 | 0.874 | 0.873 |
| | CBS | | **4.341** | **0.836** | **0.659** | **0.912** | **0.914** |
| | [7] | TOPSIS | 4.254 | 0.949 | 0.744 | 0.868 | 0.887 |
| | [7]-BC | | 4.256 | 0.946 | 0.742 | 0.869 | 0.889 |
| | CBS | | 4.126 | 1.051 | 0.858 | 0.832 | 0.878 |

Table 6 shows that incorporating Borda Count into the baseline method ([7]+BC) improves performance relative to the method [7]. Across all values of N-top, the combination of CBS similarity, KNN-based neighbor selection, and Borda Count (BC) achieves higher group satisfaction and fairness scores, together with lower RMSE-G and MAE-G than the compared method [7]. In contrast, when TOPSIS is applied for neighbor selection, CBS yields higher prediction errors and lower group satisfaction on FilmTrust. These results indicate that, for FilmTrust, KNN-based neighbor selection is more suitable than TOPSIS for producing stable and consistent group recommendation outcomes.

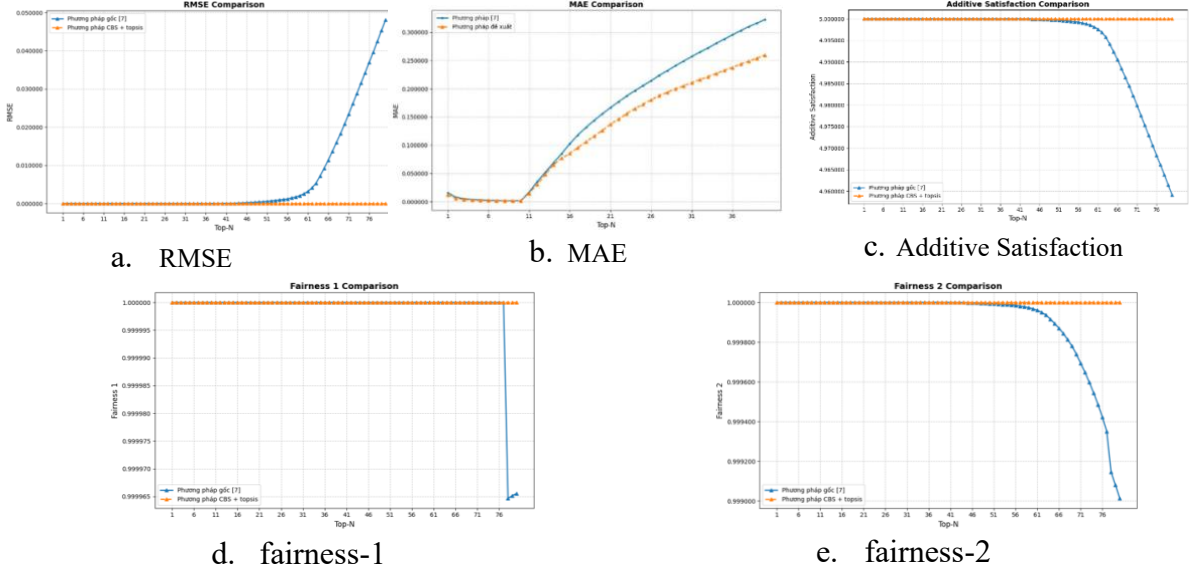

    a. RMSE      b. MAE      c. Additive Satisfaction

    d. fairness-1      e. fairness-2

**Fig. 4.** Group Satisfaction and Consensus Metrics on FilmTrust

To enable a deeper comparison with NeuMF, it is necessary to go beyond rating-based metrics. Measures such as RMSE-G, MAE-G, ADV, fairness-1, and fairness-2 are all computed based on predicted individual ratings. Consequently, additional measures that do not rely on predicted ratings, or that use original ratings only, are needed to indirectly capture the degree of group consensus with respect to the recommended item set.

It is also important to note that, unlike individual recommendation, group recommender systems explicitly prioritize intra-group consensus. When consensus is the primary objective, excessively high novelty values may be undesirable, as they may indicate that recommended items deviate substantially from preferences commonly shared within the group. In such situations, highly novel items can reduce agreement among group members, even if they are appealing from an exploratory standpoint.

Accordingly, in group recommendation, novelty should be interpreted as a balancing factor rather than a quantity to be maximized. For reference, the classical Novelty metric is recalled below.

$$Novelty\ (I_r) = \frac{1}{|I_r|} \sum_{i \in I_r} \frac{-log_2\left(\frac{|U_i|}{|U|}\right)}{log_2(|U|)} \tag{26}$$

To address this need, in addition to the classical Novelty metric, this study introduces two new trust-aware novelty measures:

- NTC (Novelty Trust Count) – *a natural extension of traditional Novelty*, quantifying objective novelty based on how rarely an item appears among trusted users in the network.

$$NTC(I_r) = \frac{1}{|I_r|} \sum_{i \in I_r} \frac{-log_2\left(\frac{|v: v\ \in U_i\ \land \exists u\ \in U: (u,v) \in E_{trust}|}{|U|}\right)}{log_2(|U|)} \tag{27}$$

- NTR (Novelty Trust Rating) – an extended novelty measure that incorporates both original rating information and trust relationships, reflecting subjective novelty through the similarity of trusted user pairs who co-rated the same item.

$$NTR(I_r) = 1 - \frac{1}{|I_r|} \sum_{i \in I_r} \sum_{(u,v) \in E_{trust}\ \land\ u \in U_i\ \land\ v \in U_i} t_{u,v} * CBS(u,v) \tag{28}$$

The two trust-aware novelty metrics proposed in this work—NTC and NTR—provide a principled extension of classical Novelty by leveraging the original rating data and the observed trust relations in the underlying social graph.

NTC evaluates novelty based on how infrequently an item appears among a user's trusted neighbors. Because it relies purely on the original frequencies of interactions in the trust network, NTC captures an objective, trust-oriented notion of novelty: items seldom consumed within trusted circles are considered more exploratory, independent of global popularity.

In contrast, NTR incorporates both trust connections and rating-based similarity computed from the original user–item ratings of trusted user pairs. By examining whether trusted users who co-rated the same item express similar or divergent preferences, NTR reflects a relationship-sensitive notion of novelty: an item is considered more novel when trusted users exhibit weak agreement regarding its rating.

Together, NTC and NTR complement the classical Novelty measure by introducing two additional dimensions—trust-aware frequency and trust-aligned rating divergence—which are especially meaningful in datasets where social links co-exist with sparse explicit ratings. This enriched evaluation perspective is essential for understanding how group recommendation algorithms behave not only with respect to global popularity, but also within the trusted regions of the user community.

In group recommender systems (GRS) where intra-group consensus is prioritized, high values of Novelty, NTC, and NTR are not necessarily desirable. Excessively novel recommendations may indicate divergence from preferences commonly shared within the group, which can undermine agreement among group members. This behavior highlights a fundamental difference between GRS and individual recommender systems (RS), where higher novelty is often encouraged to promote exploration at the individual level.

The results in Table 6 show a consistent trend: incorporating trust information meaningfully enriches novelty-based evaluation on FilmTrust. While classical Novelty reflects item popularity in the global rating matrix, the two newly proposed measures—NTC and NTR—capture trust-aware novelty at different levels. NTC provides an objective extension of traditional Novelty by quantifying how infrequently recommended items appear among trusted users. NTR further extends this notion by incorporating both trust relations and rating-based similarity between trusted user pairs, offering a subjective perspective on novelty. Together, these measures provide a more suitable evaluation framework for datasets in which social trust plays a meaningful role.

An interesting observation is that CBS, despite not using trust information at all, still produces strong novelty-related behavior on FilmTrust. This characteristic will be further illustrated in Experiment 4, where the proposed CBS-based pipeline is compared with a deep learning model. Even without relying on machine-learning–based representation learning, the non-learning collaborative filtering pipeline demonstrates competitive—often stronger—group-recommendation quality on sparse data containing only users, items, and ratings.

**Experiment 4.**

This experiment evaluates group recommendation performance on the FilmTrust dataset by comparing the proposed CBS–BC-Choquet framework with a deep-learning variant in which the individual rating prediction step is replaced by NeuMF [25]. NeuMF is chosen because it represents one of the most established neural collaborative filtering models: it delivers strong predictive accuracy using only sparse userID–itemID interactions, without requiring demographic attributes or group-level features as in approaches. This makes NeuMF an appropriate representative of deep-learning techniques designed for sparse, explicit-feedback environments.

However, NeuMF does not provide a user–user similarity measure, which is required when computing the trust-aware novelty metric NTR. Therefore, when evaluating NTR, both the CBS–TOPSIS method and the NeuMF-based variant rely on the same CBS similarity matrix to ensure fairness and methodological consistency.

The experimental setup is also retained as in Experiment 3. FilmTrust ratings are normalized to the interval [1, 5]. Except for the individual rating prediction step (performed using either CBS or NeuMF), all remaining components of the pipeline—KNN-based neighbor selection, Borda Count candidate enrichment, and Choquet-based group aggregation—remain unchanged.

**Table 7. Group Satisfaction and Consensus Metrics: NeuMF vs. CBS**

| N-top | Additive Satisfaction | | RMSE-G | | MAE-G | | fairness-1 | | fairness-2 | |
|---|---|---|---|---|---|---|---|---|---|---|
| | NeuMF | CBS | NeuMF | CBS | NeuMF | CBS | NeuMF | CBS | NeuMF | CBS |
| 5 | 3.897 | **4.528** | 1.173 | **0.608** | 0.945 | **0.471** | 0.735 | **0.974** | 0.836 | **0.952** |
| 10 | 3.851 | **4.471** | 1.189 | **0.675** | 0.959 | **0.529** | 0.721 | **0.961** | 0.832 | **0.945** |
| 15 | 3.821 | **4.438** | 1.206 | **0.715** | 0.970 | **0.562** | 0.709 | **0.950** | 0.8284 | **0.939** |
| 20 | 3.805 | **4.411** | 1.205 | **0.748** | 0.970 | **0.589** | 0.701 | **0.941** | 0.830 | **0.933** |
| 25 | 3.789 | **4.389** | 1.209 | **0.776** | 0.974 | **0.611** | 0.696 | **0.933** | 0.829 | **0.927** |
| 30 | 3.773 | **4.369** | 1.216 | **0.801** | 0.979 | **0.631** | 0.689 | **0.924** | 0.828 | **0.922** |
| 35 | 3.759 | **4.353** | 1.219 | **0.820** | 0.981 | **0.647** | 0.684 | **0.917** | 0.827 | **0.918** |
| 40 | 3.749 | **4.341** | 1.221 | **0.836** | 0.982 | **0.659** | 0.679 | **0.912** | 0.828 | **0.914** |

Table 7 indicates that, across the considered top-N settings, the CBS-based approach generally tends to be associated with higher group satisfaction values. At the same time, it is observed to yield comparatively lower prediction error measures at the group level, together with higher values of the examined fairness-related metrics, when compared with NeuMF. These observed tendencies remain relatively consistent as the size of the recommended list increases.

In addition to accuracy-oriented evaluation, we assess satisfaction and fairness dimensions of group recommendations under varying top-N list sizes. As reported in Table 8, Novelty, NTC, and NTR metrics are computed for N = 5, 10, 15, 20, 25, 30, 35, 40 compare the performance of NeuMF with the proposed CBS-based approach. These results provide further insight into how each method manages diversity, consensus, and fairness in group decision-making scenarios.

**Table 8. Novelty, NTC, and NTR Metrics for NeuMF and CBS**

| N-top | Novelty | | NTC | | NTR | |
|---|---|---|---|---|---|---|
| | **NeuMF** | **CBS** | **NeuMF** | **CBS** | **NeuMF** | **CBS** |
| 5 | 0.813 | **0.591** | 0.832 | **0.211** | 0.853 | **0.648** |
| 10 | 0.809 | **0.618** | 0.835 | **0.256** | 0.849 | **0.675** |
| 15 | 0.808 | **0.631** | 0.837 | **0.275** | 0.848 | **0.688** |
| 20 | 0.809 | **0.641** | 0.841 | **0.290** | 0.850 | **0.641** |
| 25 | 0.812 | **0.648** | 0.843 | **0.314** | 0.853 | **0.706** |
| 30 | 0.815 | **0.653** | 0.845 | **0.328** | 0.856 | **0.711** |
| 35 | 0.816 | **0.661** | 0.846 | **0.343** | 0.857 | **0.719** |
| 40 | 0.818 | **0.668** | 0.847 | **0.359** | 0.859 | **0.727** |

The results show that although NeuMF provides accurate individual-level predictions under sparse data conditions, its overall group-level effectiveness is lower than that of the proposed CBS–BC approach. Across all N values, the proposed method yields higher satisfaction and fairness scores, as well as consistently higher novelty values (Novelty, NTC, NTR). These findings indicate that CBS–BC offers a more balanced integration of predictive accuracy, group agreement, and novelty in group recommendation scenarios—particularly under conditions of extreme sparsity and heterogeneous user preferences.

## 5. Conclusion

This study proposes a consensus-driven group recommendation framework designed for scenarios in which only sparse userID–itemID–rating data are available, without demographic attributes, item metadata, social information, or predefined group structures. In such settings, user groups are inherently dynamic, which increases the complexity of individual rating prediction, candidate item selection, and preference aggregation toward stable group-level consensus.

From a methodological perspective, the proposed framework can be analyzed through four interrelated components.

**(1) Collaborative filtering under sparse explicit feedback.**

Collaborative filtering provides the foundation for individual rating prediction in group recommender systems. The analysis indicates that neighbor selection has a direct impact on group-level outcomes, and the effectiveness of strategies such as KNN or TOPSIS depends on the characteristics of the rating distribution. In addition, the number of selected neighbors ($k$) plays an important role in collaborative filtering. Beyond achieving favorable consensus-related metrics derived from predicted individual ratings, the choice of $k$ should also account for indirect indicators of group consensus that do not rely on predicted ratings, such as novelty. In consensus-oriented group recommendation, selecting $k$ in a way that leads to excessively high novelty values may be undesirable, as this can indicate increased preference divergence within the group rather than agreement.

The core idea of CBS similarity is to achieve a balanced integration of coverage, uncertainty, and discriminability in user–user similarity estimation. From a geometric perspective, CBS incorporates components related to the overlap or "area" of co-rated items, as well as vector-based similarity cues such as rating-pattern alignment, which can be interpreted through cosine-like normalization, vector length, and inner-product interactions. These elements capture how consistently two users evaluate shared items.

In parallel, CBS explicitly accounts for uncertainty under sparse observations by introducing factors that are inversely related to the amount of shared rating evidence, such as the number of co-rated items. When the available overlap is limited, uncertainty-aware components are emphasized to avoid overconfident similarity estimates.

To reconcile these aspects, CBS adopts an adaptive weighting mechanism with threshold-based switching. Depending on whether a dominant similarity signal falls below a predefined reliability threshold, CBS dynamically blends complementary components or preserves the dominant one. This conditional fusion allows CBS to operate robustly across different sparsity levels and rating distributions, while maintaining stable and discriminative similarity estimation for group recommendation.

It is important to note that group recommendation is inherently more complex than individual recommendation. Collaborative filtering models that achieve strong performance in individual rating prediction—reflected by low RMSE or MAE values, such as deep learning–based predictors (e.g., NeuMF)—do not necessarily guarantee improved group consensus when their predicted ratings are aggregated at the group level. In group recommender systems, highly accurate individual predictions may still amplify preference divergence among users, which can lead to reduced agreement after aggregation. This observation indicates that optimizing individual-level accuracy alone is insufficient for GRS, and that similarity construction and neighborhood design, as emphasized in the CBS framework, play a critical role in supporting consensus-oriented group recommendation.

**(2) Construction of the group candidate item set.**

Transforming individual recommendations into group recommendations requires an appropriate candidate selection mechanism. Selecting the top-N items for each group member preserves individual preference signals within the group. However, when predicted ratings concentrate near the upper bound, relying solely on these lists may limit discrimination. The integration of Borda Count supplements the candidate set with items that achieve relatively high ranks across multiple users, thereby strengthening group-level representation prior to aggregation.

**(3) Aggregation of individual ratings into group ratings.**

Group preference aggregation is performed using the Choquet integral, where user-specific weights determine the relative influence of each group member. These weights play a critical role in shaping the final group score, especially under heterogeneous participation and sparse observations. In the current framework, fuzzy capacities follow the formulation introduced in [7]. However, the learning and adaptation of user-specific capacities in sparse explicit-feedback group settings remain insufficiently explored.

**(4) Group consensus and trust-aware novelty evaluation.**

Group consensus cannot be fully assessed through rating-based metrics alone. To complement such measures, this study introduces two trust-aware extensions of novelty, namely NTC and NTR. Although trust information is not incorporated into the recommendation model itself, it is explicitly considered at the evaluation stage. NTC measures novelty based on original item interaction frequencies within the trust network, while NTR integrates trust relations with rating-based similarity computed from original ratings to assess novelty along trusted relationships. The experimental analysis demonstrates that the proposed framework effectively handles data sparsity and preference heterogeneity, consistently improving group satisfaction, consensus, and novelty, even in the absence of descriptive user or group information. At the same time, the results indicate that excessively high novelty values may conflict with consensus-oriented objectives in group recommendation.

**(5) Limitations**

Despite its effectiveness, CBS-based collaborative filtering exhibits certain limitations when applied to datasets characterized by high sparsity and strongly skewed rating distributions. Under such conditions, predicted ratings may concentrate near the upper bound of the rating scale, leading to top-N recommendation lists that are dominated by high-valued items and reduced differentiation among candidates.

In addition, the current CBS formulation relies on predefined parameters, including weighting coefficients and threshold values used in the adaptive combination of similarity components. These parameters are determined empirically rather than being automatically estimated from data, which may limit the flexibility of CBS when deployed across datasets with different characteristics.

Furthermore, the Choquet aggregation stage still relies on the original fuzzy capacity formulation introduced in [7], whose user-weight learning mechanism has not yet been fully explored in sparse explicit-feedback group recommendation contexts.

Future research directions will therefore focus on (i) developing learning-based approaches to estimate user-specific fuzzy capacities for more adaptive Choquet aggregation, and (ii) enhancing collaborative filtering by integrating deep-learning predictors with CBS similarity to better balance individual prediction quality and group-level consensus

**Appendix A.**

$$KNN(u,k;S) = top-k\{v \in U | v \neq u, sim(u,v) \to max\}, \tag{A.1}$$

$$TOPSIS(u,k;S) = top-k\{v \in U | v \neq u, C(u,v) \to max\}, \tag{A.2}$$

where u $\in U$, k is a positive integer, S $\in SU$, is a similarity measure between two users,

The parameters satisfy $w_s, w_{\overline{s}}, w_U > 0$, $w_s + w_{\overline{s}} + w_U = 1$, $\overline{S}(u,v) = \max(1-S(u,v)-U(u,v),0)$,

$U(u,v) = \frac{W}{W+|I_u \cap I_v|}$, where W is a positive (W = 2 in our experiments),

$$D^+(u,v;S) = \sqrt{w_s(S(u,v)-1)^2 + w_U U(u,v)^2 + w_{\overline{s}}\overline{S}(u,v)^2},$$

$$D^-(u,v;S) = \sqrt{w_s S(u,v)^2 + w_U(U(u,v)-1)^2 + w_{\overline{s}}(\overline{S}(u,v)-1)^2},$$

$$C(u,v;S) = \frac{D^-(u,v;S)}{D^+(u,v;S) + D^-(u,v;S)}$$

**Appendix B.**

The global weight assigned to each user $u$ is defined as:

$$\omega_u = \frac{|I_u|}{|I|} \tag{B.1}$$

where $|I_u|$ is the number of items rated by user $u$, and $|I|$ is the total number of items in the rating matrix.

For a given group $g$, the normalized weight of user $u \in g$ is given by:

$$\omega_{u,g} = \begin{cases} \dfrac{\omega_u}{\sum_{u \in g} \omega_u}, & \text{if } \sum_{u \in g} \omega_u < 1 \\ \omega_u, & \text{otherwise.} \end{cases} \quad \text{(B.2)}$$

For every non-empty subset $A \subseteq g$, its fuzzy capacity is computed as:

$$C(A; g) = \sum_{u \in A} \omega_{u,g} + \sum_{u \in A} \frac{|\{i \in I_u : r_{u,i} > \tilde{r}_u\}| - |\{i \in I_u : r_{u,i} < \tilde{r}_u\}|}{|I_u|} \quad \text{(B.3)}$$

Here, $\tilde{r}_u$ denotes the median rating of user $u$.

Let the users in group $g = \{u_0, u_1, \ldots, u_{|g|-1}\}$ be reordered according to their ratings for item $i$: $r_{u_{j_0}, i} \leq r_{u_{j_1}, i} \leq \cdots \leq r_{u_{j_{|g|-1}}, i}$.

The Choquet integral of item $i$ with respect to group $g$ is then given by:

$$C_g(i; g) = r_{u_{j_0}} + \sum_{k=1}^{|g|-1} \left( r_{u_{j_k}} - r_{u_{j_{k-1}}} \right) C(\{u_{j_k}, u_{j_{k+1}}, \ldots, u_{j_{|g|-1}}\}; g) \quad \text{(B4)}$$

The fuzzy capacity satisfies the following:

$$\begin{cases} C(\emptyset; g) = 0 \\ C(\emptyset; g) = 1 \\ C(A; g) = \min(C(A), 1) \end{cases} \quad \text{(B.5)}$$

A restatement of the Choquet integral is:

$$C_g(i; g) = r_{u_{j_0}} + \sum_{k=1}^{|g|-1} (r_{u_{j_k}} - r_{u_{j_{k-1}}}) C(\{u_{j_k}, \ldots, u_{j_{|g|-1}}\}; g) \quad \text{(B.6)}$$

with ordering: $r_{u_{j_0}, i} \leq r_{u_{j_1}, i} \leq \cdots \leq r_{u_{j_{|g|-1}}, i}$.